\documentclass[journal]{IEEEtran}
\usepackage{cite}
\usepackage{graphicx}

\usepackage{amsmath}

\begin{document}

\title{Analysis of coupling
losses in \\ multifilamentary untwisted BSCCO/Ag tapes \\ through
a.c. susceptibility measurements}

\author{D. Zola, F. G\"om\"ory, M. Polichetti, F. Strycek, J. Souc, P.
Kov\'ak and S. Pace
\thanks{Manuscript received 4 October
2004. This work  is partially supported by the European Commision
project ENK6-CT-2002-80658 ASTRA.}
\thanks{Danilo Zola is with
Supermat, INFM Regional Laboratory and Department of Physics "E.
R. Caianiello", University of Salerno, via S. Allende I-84081
Baronissi (Salerno) Italy, e:mail zoldan@sa.infn.it, Tel. +39 089
965369, FAX : +39 089 965275.}
\thanks{Fedor G\"om\"ory is with
Institute of Electrical Engineering, Slovak Academy of Sciences,
Dubravska Cesta 9 842 39 Bratislava, Slovakia.}
\thanks{Massimiliano Polichetti is with Supermat, INFM Regional Laboratory
and Department of Physics "E. R. Caianiello", University of
Salerno, Baronissi (Salerno) Italy.}
\thanks{Franticek Strycek is with
Institute of Electrical Engineering, Slovak Academy of Sciences,
Bratislava, Slovakia.}
\thanks{Jano Souc is with Institute of
Electrical Engineering, Slovak Academy of Sciences, Bratislava,
Slovakia.}
\thanks{Pavol Kovac is with Institute of
Electrical Engineering, Slovak Academy of Sciences, Bratislava,
Slovakia.}
\thanks{Sandro Pace is with Supermat, INFM Regional
Laboratory and Department of Physics "E. R. Caianiello",
University of Salerno Baronissi (Salerno) Italy.}}



%
\maketitle

\bibliographystyle{IEEEtran}

\begin{abstract}
Losses as function of the a.c. magnetic field amplitude ($B_0$)
were measured at 77 K in untwisted BSCCO(2223)/Ag tapes, at
different frequencies, by measuring the imaginary part of the a.c.
susceptibility. In particular, loss measurements were performed in
the portions of the same tape, obtained by cutting it in pieces
with different length, starting from around 12 cm down to 1 cm.
The results show that the measured losses depend on the sample
length but this observed behaviour is not always due to the
coupling mechanism among the filaments. In this work we discuss
the observed experimental behaviour of different typology of tapes
by analysing data comparing them with analytical models in order
to fully characterize the tapes with regard to the coupling
mechanism.
\end{abstract}



%
\IEEEpeerreviewmaketitle

\section{Introduction}
\PARstart{I}{n} A.C. electrical devices make by superconductors
cables is necessary to reduce the a.c. losses arising by the
hysteretic, coupling and eddy
mechanisms\cite{Carr83,SST10(1997)733,Clem}. Tapes or wires
usually contains many superconducting filaments because in this
way the hysteretic losses are reduced. These filaments in BSCCO
tapes, are embedded in metallic matrix, usually in silver or
silver alloy for improving the thermal stability and the
mechanical properties of tapes. On the other hand, to cut down the
coupling losses, it is essential to reduce the area of induced
flux e.g. by twisting the filaments or increasing the matrix
resistivity or manufacturing artificial resistive barriers around
the filaments. Nevertheless, the efforts turned towards a reducing
of losses in a tape can be compromised by the intergrowths and
bridging which can decrease the effective resistivity of the
matrix or electrically connect the
superconducting filaments\cite{IEEE11(2001)3557,SST14(2001)245}.\\
\indent The coupling loss  per unit volume and per cycle ($Q_c$)
if the a.c. magnetic field amplitude $B_0 \ll B_p$,which is the
full penetration field of the tape, is given
by\cite{Cryo22(1982)3,SST17(2004)501}:
\begin{equation}\label{eq1}
  Q_c = \frac{ B_0^2}{2\mu_0}\left[2\pi\chi_0
  \left(\frac{\omega\tau}{1+\omega^2\tau^2}\right)\right]
\end{equation}
where $\tau$ is so called time constant, $\chi_0$ is a constant
depending on geometry related to demagnetization factor and
$\omega$ is the angular frequency.  The Eq. \ref{eq1} shows as the
coupling loss depends on frequency and on $B_0$ square whereas the
hysteretic loss depends on $B_0^3$\cite{RMP36(1964)31,Carr83} and
it has a very slight dependence on frequency. The different
behaviour on frequency and on $B_0$ can be employed to
discriminate the different loss mechanism. The Equation \ref{eq1}
has a maximum for $\omega \tau = 1$ which can be experimentally so
estimated by finding the frequency ($\nu_m$) where the maximum
occurs in loss measurements performed at low
 $B_0$\cite{Cryo17(1977)613,Cryo17(1977)616,PC233(1994)423}.\\
At the same time, loss are linked to the imaginary part of the
first harmonic of the a.c. susceptibility
$\chi''$~\cite{Clem,PRB61(2000)6413,SST17(2004)501} according to:
\begin{equation}\label{eq2}
  Q = \pi\chi''\chi_0B_0^2/\mu_0
\end{equation}
 Both Q and $\tau$ can be also evaluated by a.c. susceptibility
 measurements also because $\chi_0$ can be experimentally
 measured\cite{SST13(2000)1327}
 Moreover, in untwisted BSCCO tapes, $\tau$ is given by:
\begin{equation}\label{eq3}
  \tau =\frac{\ell^{2}\mu_0}{\pi^{2}\chi_{0}\rho_{m}}
\end{equation}
and therefore the time constant depends on the square of the
length ($\ell$) , on the effective resistivity ($\rho_m$)and on
the tape geometry factor
($\chi_0$)\cite{PRB61(2000)6413,SST13(2000)1327}) The effective
resistivity generally differs from the bare matrix resistivity,
because it depends on the tape geometry and on the particular
arrangement of the superconducting
filaments\cite{SST13(2000)1101} and by evaluating $\tau$ the effective resistivity
can be, in this way, estimated\cite{SST17(2004)501}.\\
\indent In this work, the coupling losses have been analysed in
commercial multifilamentary untwisted tape and compared with tapes
with a geometry resembling just two filaments separated by a
metallic matrix. In this last case, each "filament" consists in
dense stack of extremely flat filaments.  We have measured by a.c.
susceptibility technique, the losses as function of frequency when
the external magnetic field is perpendicular to the broad face of
the sample.  In order to study the effect of the sample length on
the losses, the measurements were repeated for the same tape, cut
several times in shorter and shorter pieces.  Since we expect that
the hysteresis loss has not to vary as the length is changed
(whereas the coupling loss depends strongly on the sample length),
we want to characterize the quality of the tapes from bridging
aspects. Finally, the measurement have been performed in a
frequency range that the eddy loss have been estimated negligible.
\section{\label{sec1}Experimental}
The a.c. losses were measured by using an a.c. susceptometer with
a system of coils suitable for measurements on sample with length
up to 12~cm  An electromagnet produces an a.c. magnetic field with
$B_0$ up to 50~mT, with a field homogeneity within $1\%$ on a 8~cm
length and $2\%$ on 12~cm.  The a.c. field induces a voltage in
two racetrack-shaped coils: the pick-up coil, which is very close
to the sample surface, while the null coil is 1~cm apart. Since
the pick-up coil and the null coil are not perfectly identical, a
variable compensation system is also used. The a.c. magnet and
both the coils are placed in a reinforced plastic cryostat, so no
eddy currents are induced in the cryostat walls. The system is
cooled by liquid nitrogen and all the measurements have been
performed at 77~K. Measurements have been
 performed in the frequency range from 1~Hz to 1000~Hz in the
field amplitude ranging from 0.05~mT to 45~mT. \\
A.C. susceptibility measurements have been performed on two
bi-columnar tapes of around 61 mm.  The first (named in the
following "Ag tape" ) was prepared with 16 filaments in pure
silver matrix with a stack of 8 filaments for each column
separated by about 0.3~mm of pure silver. The external sheath is
also made with the same material.The geometry of the second tape
(Ag/Mg tape) is similar to that of Ag tape, but the number of
filaments is 15 and therefore there are 8 filaments in one column
and 7 in the other. The metallic sheet between filaments is a
Ag/Mg(0.4\%) alloy and the matrix which embeds the whole
filamentary zone is a Ag/Mg(0.4\%)/Ni(0.22\%) alloy. \indent
Moreover, other a.c. susceptibility measurements  have been
performed on commercial tapes, manufactured by Australian
Superconductor (AUS) and by Nordic Superconductor Technologies
(NST), whose we have few technical data. The NST samples have 65
filaments embedded in unknown matrix (probably Ag-Mg alloy) and
the cross section dimensions are 3.2 mm $\times $ 0.30~mm. We do
not know the fill factor of the tapes neither the critical
current. The Australian tapes have 37 filaments and a critical
current of 36-38~A. The cross section is 2.96 mm $\times$ 0.33~mm,
the fill factor is unknown and
 the metallic matrix is probably in Ag/Mg alloy. \\
In all tape the geometrical factor $\chi_0$ has been measured
according the technique reported in Ref. \cite{SST13(2000)1327}. I
measured value of $\chi_0$ are respectively 8.9 in Ag tape, 8.8 in
Ag/Mg tape whereas it is 3.6 in NST tape and 5.1 in AUS tape.

\section{\label{sec2}Frequency and $B_0$ dependence of losses in bicolumnar
tape}
\begin{figure}[htb]
\includegraphics[width=7.5 cm, clip]{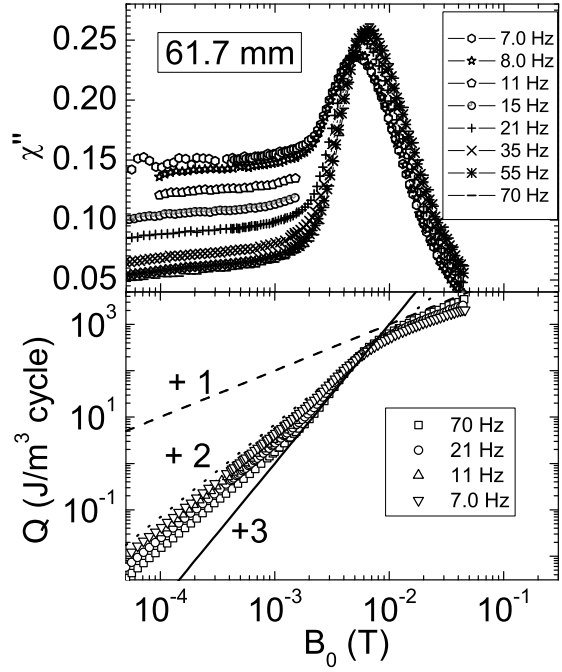}
\caption{$B_0$ dependence of $\chi''$(upper graph)and of the loss
(lower graph)in log-log scale as measured in Ag tape of 61.7 mm.
The lines shown the expected dependence, for $B_0 < B_{0,max}$, of
the hysteretic loss (continuous line) and of the coupling loss
(dotted line). The dashed lines is the expected behaviour of both
losses when $B_0
> B_{0,max}$}\label{fig1}
\end{figure}
\begin{figure}[htb]
\includegraphics[width=8 cm,  clip]{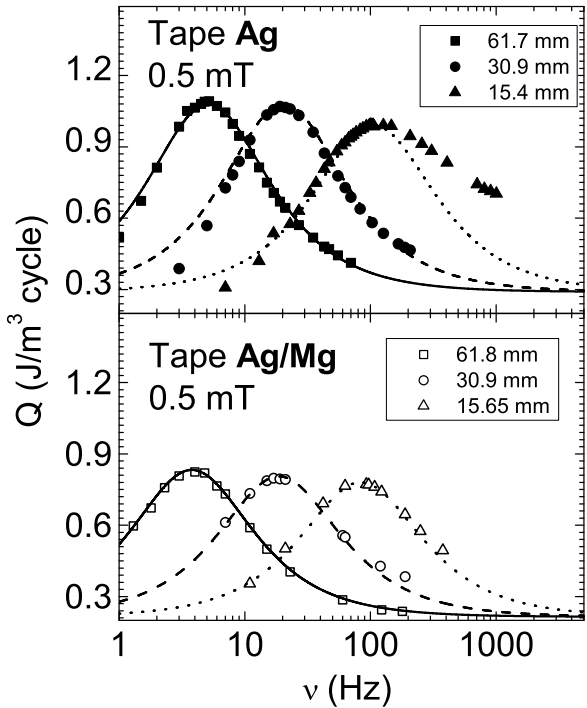}
\caption{Frequency dependence of the losses of Ag tape (upper
graph) and Ag/Mg tape (lower graph) as measured  in samples of
different lengths. The lines are the fit obtained by using the
(\ref{eq4})}\label{fig2}
\end{figure}

Losses have been measured as function of $B_0$, in several pieces
with different lengths, cut from our original Ag and Ag/Mg tapes.
In the upper graph of the Fig. \ref{fig1}, the $\chi''(B_0)$ is
shown as measured on a Ag tape of 61.7 mm  at different
frequencies. For $B_0 < B_{0,max}$(which is the field value where
$\chi''$ has a maximum and it is around two time the full
penetration field ($B_p$)of tape), the susceptibility has a large
frequency dependence. At low field amplitudes, $\chi'' $ is nearly
constant and the slight dependence on $B_0$ is probably due to the
hysteretic contribution of the superconducting columns. However,
as shown in the lower graph of the same figure, the loss densities
have a quadratic slope as expected if the coupling mechanism leads
through losses. At high field, the imaginary part and the $Q$ take
a behaviour similar to the hysteretic one due to a full coupling
of the two superconducting columns\cite{Cryo17(1977)616,Cryo22(1982)3}.\\
\indent Losses have been investigated as function of the frequency
at a fixed field amplitude much lower than $B_{0,max}$. As
reported in figure~\ref{fig2}, the losses exhibit a maximum that
shifts towards higher frequencies as the samples length decreases.
The experimental data  have been fitted by
\begin{equation}\label{eq4}
  Q_{fit}(\omega)=\alpha\frac{\omega \tau}{1+\omega^2\tau^2}+\beta
  \end{equation}
where $\beta$ is added for considering the hysteretic contribution
of the two columns. The experimental data are well fitted by the
(\ref{eq4}) with apart the experimental data measured in the
sample 15.4 mm long of the Ag tape.
The $\tau$ values, obtained from the fits, are reported in
table~\ref{tab1}and by using the (\ref{eq3}), the effective
resistivity of the metallic matrix has been determined. Since the
resistivity of the silver is 4 times smaller than that of the
Ag/Mg alloy and since the structure of the two tapes is similar,
it is surprising to see that the experimental data reported in
table \ref{tab1} demonstrate that Ag tape has a lower effective
resistivity than the Ag/Mg tape. These results can be understand
knowing that high density of inter-growths is quite
common tapes with matrix in Ag/Mg alloy\cite{SST14(2001)245}.\\
\begin{table}[tbp]
\renewcommand{\arraystretch}{1.3}
\caption{\label{tab1}Values of the quantities $\tau$,  and
$\rho_{eff}$ as determined from the experimental data for all the
considered samples.}
\centering
\begin{tabular}{c c c c }
\hline \hline
  & $\ell$ (mm)  &  $\tau$ (ms)  &  $\rho_{eff}$
($\mu\Omega$~cm)\\
\hline
Ag tape & \\
&  61.7 &  31 & 0.176 \\
& 30.9 & 7.8 &  0.175 \\
& 15.5 & 1.75  & 0.192 \\
& 6.5 & 0.32 &  0.172 \\
\hline
Ag/Mg tape &  \\
& 61.8  & 42 &  0.132 \\
& 30.9 & 8.9 &  0.155 \\
& 15.6  & 1.9 & 0.182\\
\hline \hline
\end{tabular}
\end{table}
\section{\label{sec3}Coupling losses in commercial tapes}
\begin{figure}[tbh]
\centering
\includegraphics[,width= 7.5cm,clip]{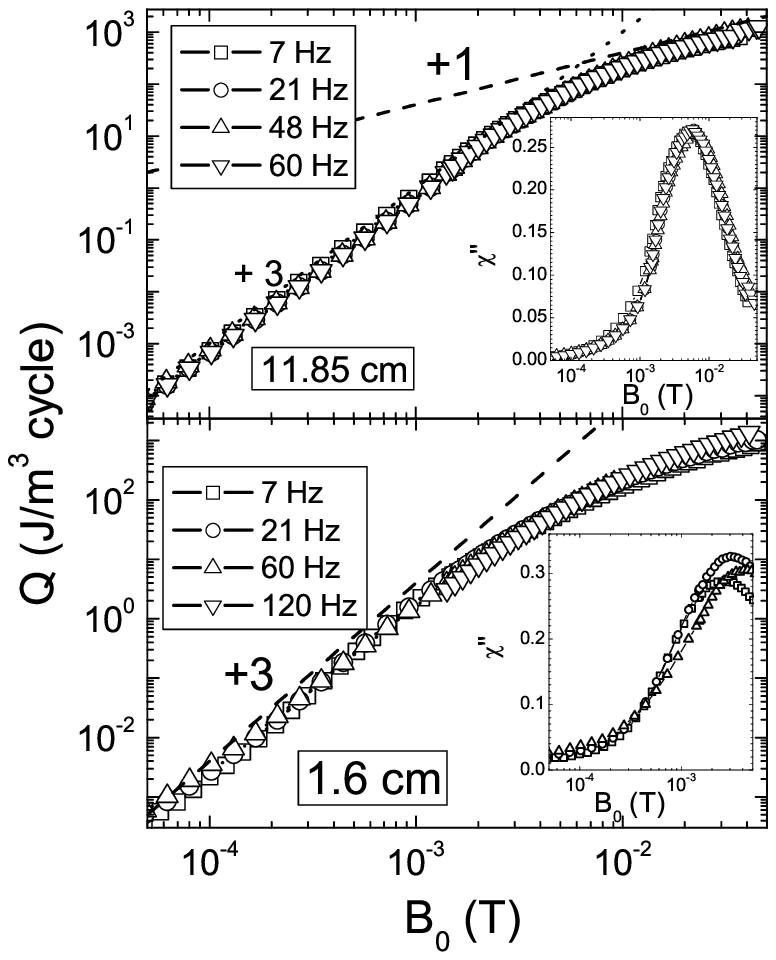}
 \caption{Losses  as function of the magnetic field amplitudes (in log-log scale) estimated on
  a NST multifilamentary tape cut in pieces of different length, measured
   at different frequencies. In the inset, the $\chi''(B_0)$ curves are shown, from which
   the losses have been estimated.}\label{fignst}
\end{figure}
\begin{figure}[tbh]
\centering
\includegraphics[,width= 7.5cm,clip]{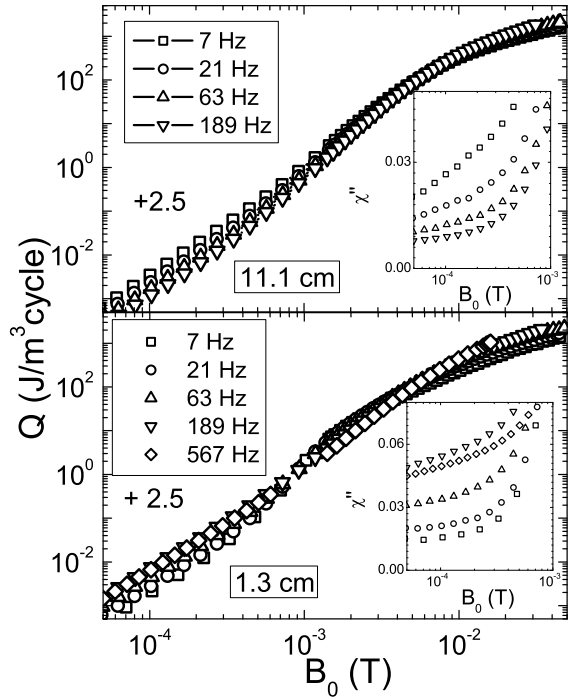}
 \caption{Losses  as function of the magnetic field amplitudes (in log-log scale) estimated on
  a AUS multifilamentary tape cut in pieces of different length, measured
   at different frequencies. In the inset, the $\chi''(B_0)$ curves are shown, from which
   the losses have been estimated.}\label{figaus}
\end{figure}
In the previous section, we have shown as by mean a.c.
susceptibility measurements, the bi-columnar tapes have been fully
characterized, in particular evaluating the effective resistivity
of the metallic matrix. On the same time, we have also verified
that the classical model on coupling losses works very well also
if this is used to analyse the coupling mechanism in BSCCO/Ag
tapes. We want to extend
this analysis to commercial tapes which are the serious candidate for applications.\\
\indent In the figures \ref{fignst} and \ref{figaus} the losses
measured on NST tapes and AUS tapes of different length are shown.
Looking at the Fig. \ref{fignst}, we can observe in  upper and
lower graph that the losses have a cubic dependence at low field
whereas the slope becomes +1 at high field. This behaviour
behaviour is observed on the sample 11.85 cm long as in the sample
of 1.6 cm and equivalent results have been found at intermediate
lengths. At the same time, the imaginary part of the a.c.
susceptibility has not any frequency dependence. The analysis
performed on these tapes lead us to conclude that in NST tape the
coupling are completely suppressed for effect of a very strong
bridging among the filaments that lead to a behaviour very similar
to a monofilamentary tape with hysteretic loss only.\\
\indent In AUS tapes we can observe a more interesting behaviour.
The losses have been measured on samples with length starting from
11.1 cm down to 1.3. At low field, the slope of the losses in
log-log scale is close to 2.5 as on sample 11.1 cm long as on a
piece of only 1.3 cm. The value of the slope confirms that the
coupling mechanism competes with hysteretic one for every length.
In the two inset of the figure, the imaginary part of the AUS
tapes as measured in a piece of 11.1~cm and 1.3~cm show a some
frequency dependence. Similar measurements,no shown, have been
performed on a sample 5.5~cm long and they have an equivalent
behaviour. In particular in the two inset of Fig. \ref{figaus} we
can observe the low field region of the a.c susceptibility
measured at different frequencies. In the sample 11.1 cm long, the
$\chi''$ decreases as the frequency increases and this could
meaning that the maximum of the coupling losses is at lower
frequency. As shown in previous section, cutting the sample this
maximum shift at higher frequency. In the 5.5 cm sample the
$\chi''$ seems to have a maximum between 7 Hz and 21 Hz whereas in
the sample 1.3 cm long this maximum seems to be at frequency
around 189 Hz with a saturation measured up to 567 Hz. The
frequency trend of AUS tape does not resemble exactly the quasi
ideal behaviour of the bi-columnar tapes. On the other hand, we
know that in pure metal the value of $\chi''$ when $\omega\tau =
1$ is 0.38 whereas it is, in general,  smaller in a filamentary
superconductor surrounded by a metal. For example in the
bicolumnar tapes, where the coupling mechanism is dominant, at 0.1
mT the maximum value of the susceptibility is around 0.15 which is
3 times larger than the value measured on AUS tapes at the same
field.  All these considerations lead us to conclude that in AUS
tapes a part of tape is strong bridged probably in the more
densely packed region but an other part the bridge is not so
strong and a coupling mechanism can arise. Nevertheless the large
hysteretic terms is comparable with the coupling loss and we
cannot fully characterize the samples. However, through this
investigation we have acquired important informations on the a.c.
behaviour of tapes potentially employed in the realization of
applications investigating deeply the filamentary nature of the
BSCCO tapes.

\section{\label{sec4}Conclusions}
In this work we have studied the a.c. coupling losses on two
different sets of tapes, bicolumnar BSCCO tapes and  commercial
tapes. A.C. susceptibility and losses, measured as function of the
magnetic field amplitude and frequency, show that the coupling
losses dominate in bi-columnar tapes over the hysteretic losses of
the single filaments. The experimental effective resistivity in Ag
tape has a higher value in comparison with Ag/Mg tape which has a
$\rho_{eff}$ lower than expected. This results are explained with
the presence of intergrowths in Ag/Mg tapes and this  result
suggests that in this kind of tapes, the enhancement of the
effective matrix resistivity does not reduce automatically the
coupling losses. In NST tapes we have found a very strong bridging
which lead to suppress completely the coupling mechanism and the
tape works like a monofilamentary tape. Finally in AUS tapes the
coupling mechanisms competes with a large hysteretic component,
probably due to a strong bridging in a significatively part of the
tape. Our work show as through a.c. susceptibility it is possible
to investigate the quality of BSCCO/Ag multifilamentary tapes.

\bibliography{1085}

\end{document}